\documentclass[letter,paper]{emulateapj}
\newcommand{\Ks}{$K_{s}$}
\newcommand{\ha}{H$\alpha$}
\newcommand{\hb}{H$\beta$}
\newcommand{\sii}{[S\,{\small II}]}
\newcommand{\feii}{[Fe\,{\small II}]}
\newcommand{\um}{$\mu$m}

\newcommand{\AV}{$A_{V}$}

\shorttitle{The Reddening towards Cas A}

\begin{document}




\title{The Reddening Towards Cassiopeia A's Supernova: Constraining the $^{56}$Ni Yield}

\author{Kristoffer A. Eriksen\altaffilmark{1}, David Arnett, Donald W. McCarthy}
\affil{Steward Observatory, University of Arizona, Tucson, AZ 85721}

\and

\author{Patrick Young}
\affil{School of Earth and Space Exploration, Arizona State University, Tempe, AZ, 85287}

\altaffiltext{1}{Visiting Astronomer, Kitt Peak National Observatory.
KPNO is operated by AURA, Inc.\ under contract to the National Science
Foundation.}

\begin{abstract}
We present new reddening measurements towards the young supernova remnant Cassiopeia A,
using two techniques not previously applied to this object. Our observations of the
near-infrared \feii\ $1.257\mu m$\ and $1.644\mu m$\ lines show the extinction to
be highly variable across the remnant, increasing towards the west and the south, consistent 
with previous radio and X-ray observations. 
While the absolute value of \AV\ as determined by the
\feii\ lines is uncertain due to conflicting calculations and observations
of their intrinsic flux ratio, parts of the remnant without previous optical
measurements show comparatively higher reddening.
We find $A_{V} = 6.2 \pm 0.6$\ from the broadband shape of 
the infrared synchrotron emission of a knot within 13\arcsec\ of the expansion center.
Given this reddening, the apparent faintness of the nascent supernova, and iron mass
constraints from X-ray observations, we estimate an ejected mass of $^{56}$Ni of
$0.058 - 0.16 M_{\sun}$. Taken with $\gamma$-ray observations of the $^{44}$Ti decay
chain, this nickel mass is broadly consistent with the solar $^{44}$Ca/$^{56}$Fe ratio.

\end{abstract}

\keywords{supernova remnants --- supernovae: individual (SN1680) --- dust, extinction --- nuclear reactions, nucleosynthesis, abundances }

\section{Introduction}

Cassiopeia A is the remnant of one of the most recent known supernova in the
Galaxy. It is the prototype of the ``oxygen-rich'' class of supernova remnants (OSNRs),
whose optical spectra are dominated by highly Doppler shifted lines of
oxygen and other advanced burning products, but are nearly devoid of hydrogen and helium.
Because these remnants have not yet swept up significant interstellar material,
their abundance pattern reflects that of their progenitor star and supernova.
As such, the OSNRs, and Cas A in particular, allow unparalleled access to the physics
of massive stars and supernova, and should provide tight constraints on models of
nucleosynthesis, stellar mixing, supernova physics, and ISM enrichment.

Given its distance ($3.4^{+0.3}_{-0.1}$\ kpc, Reed et al. 1995), 
circumpolar location in the northern sky, and the most quoted reddening
towards Cas A ($A_{V} \sim 5$, Hurford \& Fesen 1996, hereafter HF96), 
it is surprising that its nascent supernova was largely unreported. 
SN1604 (Kepler's SN),
SN1572 (Tycho's SN), SN1054 (the Crab Nebula) and SN1006 were all widely observed
and recorded by contemporary astronomers in Europe, the Far East, or both.
Cas A was not. While Flamsteed may have seen the
event as a 6th magnitude optical transient in 1680 (Ashworth 1980, though 
Kamper 1980 provides a dissenting opinion),
it was certainly less 
remarkable than other historical Galactic supernovae. Why?
Was the Cas A supernova an unusual, under-luminous
event? Were contemporary observers unlucky, and the supernova at peak brightness
was highest in the northern sky during daylight? Is the extinction towards Cas A
greater than previously assumed?

Aside from this historical oddity, the apparent faintness of Cas A's outburst
presents an astrophysical conundrum.  The peak luminosity and
characteristic light curve of supernovae are driven by energy injected into
the expanding stellar debris by the radioactive decay of $^{56}$Ni and its
daughter nucleus $^{56}$Co. Thus, the peak brightness is proportional to the ejected
mass of $^{56}$Ni. Since the complete explosive silicon burning that produces radioactive
nickel takes place in the deepest layers of the supernova, the nickel yield is
intimately linked with the explosion mechanism, energy, mass cut, and fall-back fraction.
Its diagnostic power is further enhanced when compared to a supernova's $^{44}$Ti yield.
This trace isotope is produced by ``$\alpha$-rich freeze-out,'' a process of incomplete
silicon burning that dominates in regions of higher entropy (i.e. lower density). 
The high $^{44}$Ti abundance,
as determined by $\gamma$-radiation observations of its decay chain 
(Iyudin et al. 1994, Vink et al. 2001)
relative to the implied $^{56}$Ni mass inferred from Cas A's apparently faint outburst,
are inconsistent with yields expected from symmetric explosion calculations
(Timmes et al. 1996). This is generally assumed to be evidence for asymmetry in
the supernova (e.g. Nagataki et al. 1998). Therefore, tighter measurements of 
$^{44}$Ti/$^{56}$Ni should provide helpful constraints on advanced supernova explosion
calculations.
Furthermore, $^{44}$Ti appears to be produced exclusively in core collapse
and sub-Chandrasekhar mass SNIa (Timmes et al. 1996). 
A measurement of radioactive titanium and nickel
in Cas A may inform Galactic chemical evolution and stellar population
synthesis models in determining the relative importance of these two types of supernovae
in producing the observed solar abundance ratio of $^{44}$Ca/$^{56}$Fe, the two 
stable daughter products of these nuclei.

Thus a proper inventory of $^{56}$Ni or, equivalently $^{56}$Fe, in Cas A is desirable.
Willingale et al. (2003), using XMM-{\it Newton} X-ray observations, find an X-ray emitting
iron mass (presumably dominated by $^{56}$Fe) of $0.058 M_{\sun}$. Because the cooling
time of the X-ray emitting plasma is generally longer than the age of the supernova 
remnant, this is, to a good approximation, the sum total of iron that has been heated to
X-ray emitting temperatures upon passage through the reverse shock. However, it does
not include denser material that retained enough electrons to cool efficiently through
EUV, optical, and infrared emission. The cooling times for these plasmas can be weeks or 
months, making a full inventory of gas of these densities at reverse
shock passage difficult. Similarly, material that has not yet encountered the reverse shock
is nearly invisible.

Here we take a different approach and attempt to bound the allowed total
mass of ejected $^{56}$Ni. Given that Cas A's SN was apparently unremarkable
at Earth, knowledge of the extinction towards the supernova,
a distance measurement, and a limit on the apparent magnitude
place constraints on the intrinsic peak luminosity of the event.
Previous optical measurents of the reddening suffer from a selection effect, 
in that the most common method for estimating \AV\ is infeasible in regions
of higher extinction.
Existing CO (Troland et al. 1985) and X-ray (e.g. Keohane et al. 1996)
observations show the absorption to Cas A to be spatially variable,
with a general trend of increasing extinction from east to west,
though poor spatial resolution washes out structure on scales smaller
than the beam and will underestimate
\AV\ if the absorbers are clumped. Indeed, interferometric observations of
H$_{2}$CO absorption towards Cas A (Reynoso \& Goss 2002)
show considerable clumpiness.
Moreover, radio and X-ray
measurements of the absorbing column require an assumed gas to dust ratio to convert
to optical extinction.
As such, existing estimates of the reddening span several magnitudes, most commonly
$A_{V} \sim 4-8$.
(Fesen et al. 2006 assume $A_{V} = 6-8$\ towards the
compact remnant.) 
These uncertainties have limited the accuracy of previous attempts
(e.g. Troland et al. 1985)
to use the peak brightness to constrain the properties of the supernova.
Here we present significantly
improved measurements of \AV\ using \feii\ infrared emission lines and
a new technique based on measurement of the infrared spectral index of the synchrotron
emission of Cas A.

The rest of the paper is organized as follows. 
In \S 2, we detail the new near-infrared images
and spectroscopy and the archival mid-infrared imaging we use for our measurements.
In \S 3 we address the challenges of measuring the extinction towards Cas A with emission
lines, estimate \AV\ from our \feii\ observations, and describe and present results 
from our measurement of the infrared synchrotron emission. In \S 4, we bound the ejected nickel
mass and discuss its implications. \S 5 is a summary of our results.

\section{Observations}

\subsection{Near-Infrared Imaging} 

We imaged Cas A with the PISCES (McCarthy et al. 2001) near-infrared camera
attached to the Steward Observatory
2.3-meter Bok telescope, over two different runs in summer 2002 and 2003.
The 8.\arcmin 5 circular field of view and 0.\arcsec 5 pixels at the
Bok f/9 focus allowed us to image the entire remnant in a single pointing
while adequately sampling typical Kitt Peak seeing,
which ranged from 0.\arcsec8-1.\arcsec5. All observations
were made with broadband filters from the 2MASS system.

In July 2002, we used a 3x3 dither pattern, with 60 second individual
exposures in {\it J}, and 30 second exposures in {\it H}. 
The total exposure times were 54 minutes in {\it J} and 40 minutes in
{\it H}. In July 2003, we used a similar strategy, observing
Cas A in {\it J} (45s individual, 45 minute total exposure),
{\it H} (12s individual, 35 minutes total exposure) and 
\Ks\ (10s individual, 56 minutes total exposure).

Standard image processing, including dark subtraction, flat-fielding,
distortion correction, and image stacking were accomplished with the IRAF 
\footnote{IRAF is distributed by the National Optical Astronomy 
Observatory, which is operated by the Association of Universities
 for Research in Astronomy (AURA) under cooperative agreement with the
 National Science Foundation.} 
package. For the 2002 run, flat-field
frames were constructed from median-filtered off-source dithers
interspersed with on-source pointings.
In 2003, no off-source pointings were taken, and flats were made from
source frames. This resulted in a poor quality flat-field, and an
underestimation of the flux in
low-surface brightness, high-coverage features like the diffuse synchrotron
emission observed in all bands. As such, the 2003 {\it J} and {\it H} images
were not used for flux measurements, though they were used to estimate a 
correction factor for the 2003 \Ks\ data. Magnitude calibration was achieved
by comparison with 2MASS (Skrutskie et al. 2006) stars in the observed field.

We present in Figure 1 a color $JHK_{s}$ image of Cas A from our PISCES data.
The clumpy shocked material has strong lines in the {\it J} and {\it H} bands
and appears blue-green, while the \Ks\ image is dominated by smoother
synchrotron continuum. While this non-thermal emission is detected in all bands,
it is strongest at longer wavelengths, and displays red.

\begin{figure}
\plotone{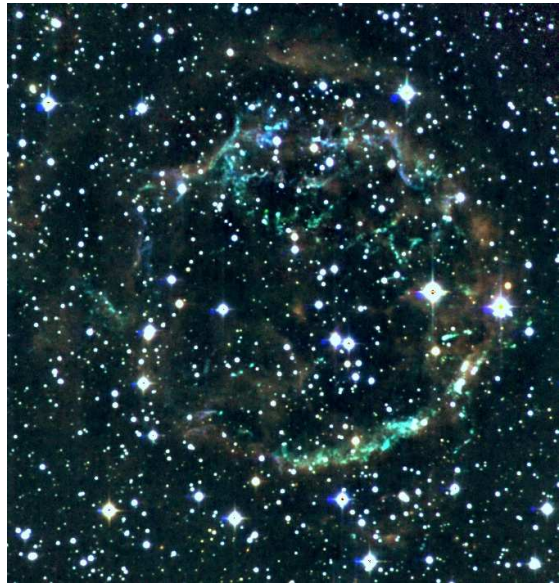}
\caption{Our color $JHK_{s}$ (red, green, blue) PISCES image of Cassiopeia A. 
The clumpy blue/green features are line emission from shock heated ejecta and circumstellar 
material, while the smoother red features are synchrotron emission. Synchrotron ``Knot N''
from Wright et al. (1999)
is clearly visible as a diffuse red feature near the center of the remnant.}
\end{figure}

\subsection{Near-Infrared Spectroscopy}

Using our {\it J} band image to target knots of interest,
we observed four slit positions on 26-27 September 2004 
with the FLAMINGOS
\footnote{FLAMINGOS was designed and constructed by the IR instrumentation
group (PI: R. Elston) at the University of Florida, Department of Astronomy,
with support from NSF grant AST97-31180 and Kitt Peak National Observatory}
multi-object spectrograph on the
KPNO 4-meter telescope. The JH grism and 6 pixel wide long-slit provided
simultaneous wavelength coverage from approximately 0.95-1.8\um\ across
a $10\arcmin \times 1.9\arcsec$\ slit on the sky, and delivered approximately
25\AA\ resolution across the J and H bands. The emission knots in Cas A
are densely packed on the sky, so it was impossible to use the standard 10\arcsec\ nod along
the slit for sky subtraction. Since the FLAMINGOS slit is approximately 
twice as long as the diameter of Cas A's bright ring,
we placed the all supernova remnant line emission at one end of the
slit, and used a 5\arcmin\ nod. While this large throw cost extra
overhead (including reacquisition of a guide star at each nod), it ensured
knots of interest would not overlap between nods. Utilizing an ``AABB''
pattern, each individual exposure was 300 seconds, for a total A+B on-source
exposure time of 4800-6600 seconds. 

Standard near-infrared spectroscopic reductions (dark subtraction, flat-fielding,
sky subtraction, distortion correction) 
were accomplished with a combination of IRAF and custom
Perl Data Language
\footnote{The Perl Data Language (PDL) has been developed by K. Glazebrook, J. Brinchmann, J. Cerney, C. DeForest, D. Hunt, T. Jenness, T. Luka, R. Schwebel, and C. Soeller and can be obtained from http://pdl.perl.org}
(Glazebrook \& Economou 1997)
scripts. For the J band, we set the wavelength
calibration with HeNeAr lamps. H band wavelength calibration was made against
the night sky lines, using VLT/NIRMOS OH line list and sky spectrum convolved to
our resolution (Rousselot et al. 2000). 
The instrumental throughput and telluric transmission
was measured with observations of the nearby G2V star HD212809, divided by
the synthetic solar spectrum included with the spectral synthesis
code SPECTRUM v2.73 (Gray \& Corbally 1994). Small corrections to the 
shape of the instrumental response were corrected with observations of the A0V
star HD240290, relative to a synthetic Vega spectrum calculated with the same
code. Finally,
the absolute flux scale was set with an observation of G191B2B, utilizing the
STIS/NICMOS fluxes (Bohlin 2007).

\subsection{{\it Spitzer} Mid-Infrared Imaging}

We used archival {\it Spitzer}/IRAC images of Cas A originally presented by
Ennis et al. (2006). While data exists for all four IRAC bands, the
broader point spread function for the longer wavelengths degrades our ability to
disentangle the synchrotron emission peaks from both the emission-line emitting 
regions and the smooth continuum background, so we limit our analysis to 
channels 1 \& 2 (3.6$\mu$m \& 4.5$\mu$m, respectively). Data products were downloaded
from the archive, processed and mosaicked using standard IRAC processing.

\section{The Extinction Towards Cas A}

\subsection{Reddening Measurements from Emission Lines}

Reddening measurements of emission-line nebula require observations of 
at least two lines of known intrinsic ratio and significantly different energies. 
With the adoption of a
reddening curve (e.g. Cardelli et al. 1989, which we use here)
and an assumption about general-to-selective extinction
($R_{V} = A_{V}/E(B-V) = 3.1$\ is the standard value for the diffuse ISM),
the observed fluxes of the selected lines gives the extinction. \ha\ and \hb\  
(and higher order lines of the Balmer series) are most often used since
they are generally among the brightest lines in nebular spectra and, in the
usually applicable case
where their emission is primarily by recombination, the intrinsic ratio is
well-known. However, these lines are problematic in Cas A for two reasons. 
First, the
fast-moving knots that dominate Cas A's optical spectra are composed of pure
metals and are devoid of hydrogen lines. Second, while the quasi-stationary flocculi
(``QSFs'', shocked pre-SN circumstellar material)
do emit in the Balmer series, the slow radiative shocks that illuminate these
knots produce \ha\ by both recombination and collisional excitation, making
the intrinsic \ha/\hb\ (Balmer decrement) ratio uncertain. A value of 3.0 is
usually assumed for radiative shocks, but this increases to slower shock
speeds as collisional excitation becomes more important. HF96 report measurements
of the Balmer decrement for two QSFs in Cas A, but regard their implied values
($A_{V} = 5.3, 6.2 \pm 0.9$) as upper limits.

Alternatively, it is possible to measure the extinction with metal line ratios. 
In the case where
collisional de-excitation is negligible, the intrinsic flux ratio of
lines originating from the same upper term depends only on their energies
and transition probabilities. Unfortunately, calculation of the Einstein
A coefficients for the forbidden lines most often observed in optical/IR 
emission-line nebula is a difficult task, and results can vary by 20\% or
more for different calculations of the same transitions. 
Moreover, if collisional de-excitation is important 
the intrinsic flux ratio depends both on the
electron density and theoretically-calculated effective collision
strengths, which may have systematic uncertainties equal to 
or greater than those of the transition probabilities. 

The best extinction measurements for Cas A to date are from HF96,
who examine the 1.03\um\ blend and $\lambda\lambda4069,4078$\AA\ doublet
of \sii, both of which originate in the $^{1}P$\ (second excited) term of
S$^{+}$.
(Collisional de-excitation of these transitions is unlikely to be important
in the Cas A FMK's, as their critical densities are of order $10^{6}$~cm$^{-3}$.)
The issue of the accuracy of the necessary atomic data is apparent in
HF96, who assume a ratio 25\% greater than had been used in an earlier study by
Searle (1971). Aside from systematics related to the atomic data, two practical
considerations limit the utility of the \sii\ diagnostic. As noted above,
the lines of interest originate from the second excited term of S$^{+}$, 
so, unlike the ubiquitous \sii\ $\lambda\lambda6717,6731$\AA\ doublet,
at lower temperatures 
it is possible to have a significant S$^{+}$\ ionization fraction with
little population of the higher term, resulting in weak lines.
More importantly, the 4070\AA\ lines are far into the blue,
and are thus highly absorbed towards Cas A. Indeed, for $A_{V} = 8$, the
highest derived extinction inferred from the radio study of 
Troland et al. (1985), $A_{4070} > 11$. HF96 only report reddenings 
in the northeastern bright ring of Cas A, where the radio-inferred extinction
is the least. This is likely a selection effect, as it becomes increasingly 
difficult to measure the faint 4070\AA\ lines in heavily absorbed regions.
Since radio studies show the reddening towards
Cas A to be patchy and variable across the remnant, in order to 
constrain more tightly the extinction at the explosion
it is highly desirable to be able to measure reddenings closer to the expansion
center or, at least, sample a more representative range of sight lines.

With the general availability
of efficient near-infrared spectrographs, it is now possible
to observe a wider selection of emission lines into regions of higher
extinction. The strong 1.257\um\ and 1.644\um\ lines of \feii\ both originate in
the $a^{4}D_{7/2}$ level of Fe$^{+}$, are bright in a range of emission-line
nebulae, and in principle should be ideal reddening
indicators for moderately absorbed sources. However, the intrinsic flux ratio
of these important lines is a matter of some debate. Two modern calculations of
the relevant transition probabilities (Nussbaumer \& Storey 1988 and 
Quinet et al. 1996) are discrepant by 30\%. Because of the relatively small
wavelength spacing of the two lines, this translates to a systematic
error in \AV\ of several magnitudes. Hartigan et al. 2004 use the Quinet ratio 
($F_{1.2575}/F_{1.644} = 1.04$) in 
their models, since it implies an \AV\ in HH111 that is consistent with optical
Balmer decrement measurements (Gredel \& Riepurth 1993). However, Nisini et al.
(2005) find that neither the Quinet nor the Nussbaumer \& Storey ratios produce
produce \AV's consistent with optical data in HH1. In light of such theoretical
uncertainty, an empirical determination is desirable. To our knowledge, there are no
published $1.257\mu m/1.644\mu m$\ flux ratios for 
relatively unabsorbed sources (e.g.s. The Cygnus
Loop or the Orion Nebula), but Smith \& Hartigan (2006, hereafter SH06) 
present high S/N
J and H band spectroscopy of P Cygni's nebula, and derive empirical Einstein
A coefficients
from their observations, using the known optical reddening. This is, of course,
dependent on their assumed extinction ($A_{V} = 1.86$, Lamers et al.  1983), itself
reliant on the accuracy of B star NLTE model atmospheres, being correct. 
However, given the
current uncertainty in the actual intrinsic ratio, we accept the SH06
measurement ($F_{1.257}/F_{1.644} = 1.49$) as the best available, and use it throughout
this work.

\begin{figure}
\plotone{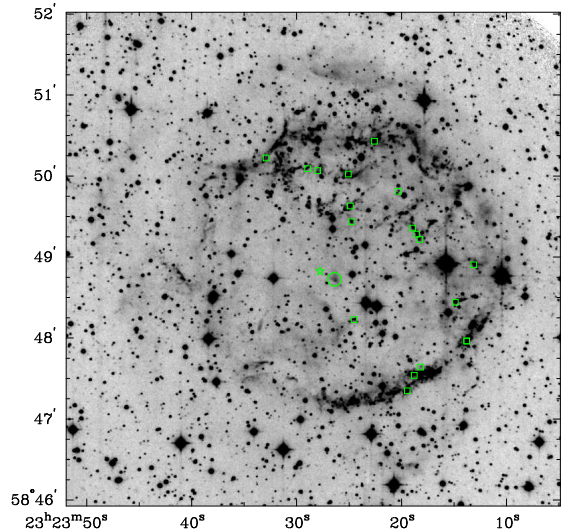}
\caption{PISCES {\it H} band image of Cas A. Our knots with successfully measured \feii\ 
ratios are marked with boxes, and synchrotron knot N is marked with a circle. 
The star symbol marks the expansion center as measured by Thorstensen et al.
(2001).}
\end{figure}

\begin{figure}
\plotone{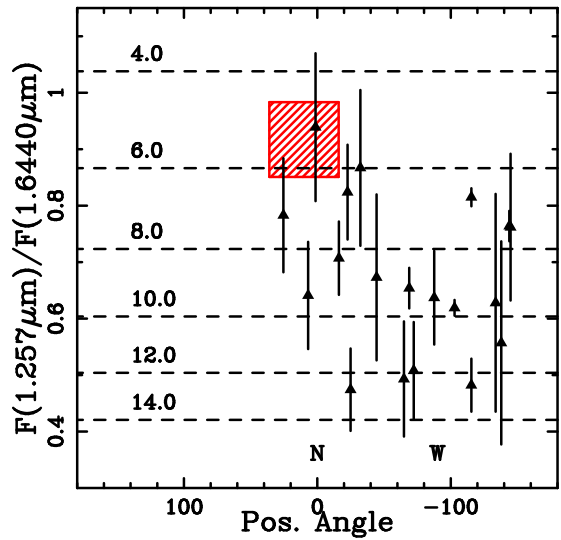}
\caption{The observed flux ratios for 19 \feii\ knots. The horizontal lines denote \AV\
assuming the empirical intrinsic flux ratio of Smith \& Hartigan (2006).
Adoption of the Nussbaumer \& Storey (1988) ratio would decrease \AV\ by more than a
magnitude. The hatched region shows the range of position angles sampled 
by HF96, and the range of reddenings from their optical \sii\ lines.}
\end{figure}

\subsection{\feii\ Reddening Measurements}

We extracted numerous spectra from our coadded observations, seeking to
maximize the signal to noise in the \feii\ lines of interest. 
In Figure 2, we mark the positions of the 19 
individual knots in which we measure the $F_{1.257}/F_{1.644}$\
ratio to better than 20\%.
Of these, 16 are fast-moving ejecta knots and 4 are QSFs,
located both in the northern bright ring 
(where HF96 made their measurements) and the more highly reddened western and southern
portions of the remnant.  In Figure 3, we show the measured flux ratio against
position angle in the remnant. While there is considerable
scatter due to Poisson noise in the lines, there is significant real variation 
on small spatial scales,
as well as a general increase in reddening from east to west. 
Both these attributes are
broadly consistent with with the radio measurements of Troland et al. (1985),
who estimated $A_{V} \sim 4-5$\ for much of the remnant, increasing
to $A_{V} \sim 8$\ in their western most observation.
However, they argue that much of the molecular gas
along the line of sight must be clumped in dark clouds smaller than their 1.1\arcmin\ 
beam, so  the extinction would be expected to be greater in places,
and vary on scales smaller than an arcminute.

Aside from the striking variation in reddening, an overall offset in the implied \AV\
as compared with the \sii\ measurements from HF96 is clear. 
While there is no direct
overlap in knots observed between the two samples, a number of our features are in the same
area of the bright northern ring as the the HF96 observations. Yet, there is a marked
difference of several magnitudes. The most likely explanation for this is the still
quite uncertain atomic data for the forbidden lines of interest.
A small portion of this error may
come from the \sii\ measurements. Our own calculation of the intrinsic \sii\ ratio using
more modern atomic rates (Keenan et al. 1993 and Keenan et al. 1996) 
suggests $F_{1 \mu m}/F_{4070} = 0.58$, a revision
down by approximately 10\% compared with HF96's adopted value. However, given the large
separation in wavelength of the two blends, this only accounts for a few tenths of a
magnitude in \AV. Rather, the likely culprit is the large uncertainty in the
intrinsic relative
strength of the two \feii\ lines. Indeed, if we were to assume the theoretical 
value of Nussbaumer \& Storey (1988) rather than the empirical ratio of SH06,
the derived \AV\ would decrease by more than a magnitude, and would bring
our least reddened values closer in line with HF96. This adjustment is problematic
though, in that it would imply essentially zero extinction to P Cygni given SH06's
observed flux ratios, in conflict with ultraviolet and optical observations. Clearly,
an empirical measurement on a less absorbed object is required to fix this distressing
situation. Until its resolution, we caution that 
absolute measurement of \AV\ with these lines,
for this SNR or any of the large number of astronomical objects 
that would benefit from NIR emission-line extinction diagnostics,
requires better atomic data.  However,
apart from an absolute measurement,
the variation in the observed flux ratio is a clear
indication of the high spatial variation of the reddening toward Cas A, and
demonstrates that an accurate estimate of the extinction of Cas A's nascent supernova
requires a measurement of \AV\ as close to the expansion center as possible. Our
ability to reliably measure this ratio across the face of the remnant, even towards
regions of comparitively high absorption, is a clear advantage over optical 
measurements.

Finally, we note the existence of two knots, one FMK and one QSF, within 45\arcsec\
of the measured expansion center of Cas A. The reddenings for these knots are 
$A_{V} = 7.4^{+2.1}_{-1.7}$\ and $A_{V} = 6.0^{+1.9}_{-1.6}$\ (statistical errors only). 
These are the closest emission line features
to the projected explosion site to date with measured reddenings.

\begin{figure}
\plotone{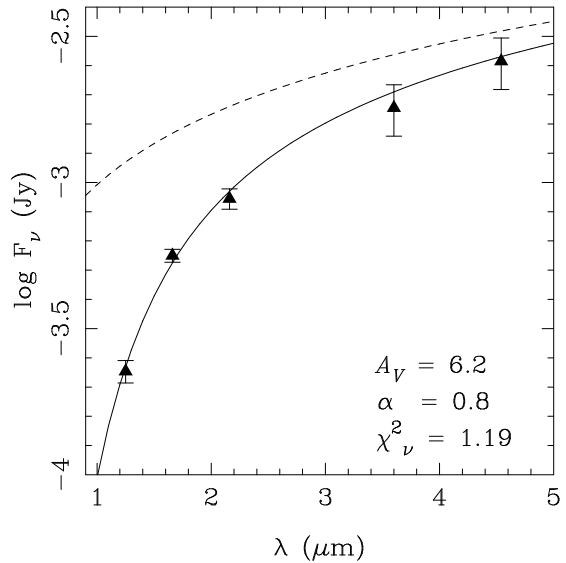}
\caption{Measured infrared flux densities for Cas A synchrotron knot N. The solid line
shows the best fit model for $A_{V} = 6.2$\ and $\alpha = 0.80$, while the dashed line
is the unabsorbed model.}
\end{figure}

\subsection{Reddening Measurement from Infrared Synchrotron Emission}

Relativistic electrons accelerated in Cas A's shocks produce copious amounts
of synchrotron emission, and place 
Cas A among the brightest sources in the centimeter and millimeter sky. 
Cas A's X-ray spectrum
has a hard, broken power-law component (Allen et al. 1997) 
in additional to its thermal plasma
signature, and {\it Chandra} has revealed a number of discrete filaments with 
featureless spectra (e.g. Hughes et al. 2000), which are generally assumed to be
synchrotron emission (though, see also Laming 2001). It is then perhaps not surprising
that synchrotron emission would be detected in the $\sim 8$ decades of energy in between.
Gerardy \& Fesen (2001) were the first to present high-quality near-infrared images
of Cas A, and while their {\it J} band image was similar in morphology to
clumpy, filamentary optical images dominated by emission lines, they noted the
striking similarity of their \Ks\ image to radio continuum maps, and postulated
they had detected infrared synchrotron emission. Jones et al. (2003, hereafter J03) 
confirmed this identification
with {\it K}-band imaging polarimetry of a small portion of the western shell.
They also found the 2\um\ flux to be brighter than expected from simple
extrapolation from the radio. Rho et al. (2003, hereafter R03) apparently confirm the 
``concave-up'' nature of the Cas A synchrotron spectrum, though since
their measurement is summed over the entire remnant, systematics like incomplete
star subtraction, contamination from line-emission, spectral index variation,
and large-scale flat field erros are likely sources of systematic error.
Indeed, Wright et al. (1999, hereafter W99) show significant variation
of spectral indices for several knots in the millimeter-wave regime.

Departures from a broadband power-law spectrum, either concave up or down, can 
be produced by a variety of plasma effects (e.g.s Eichler 1984, Reynolds 1998),
and the detailed broadband shape of Cas A's non-thermal SED is still an open question.
Of course, at optical through mid-infrared wavelengths, the situation is complicated
by the substantial extinction along our sight line to the remnant. (For $A_{V} = 4.5$\
and a standard reddening law, $A_{4.5\mu m} = 0.16$\ and is obviously greater for 
shorter wavelengths.) Here, we reverse the question:
we assume the departure from a power-law is small over our limited range of 
energies, and determine \AV\ by fitting for the observed fall-off towards
shorter wavelengths.

Practical measurement of the infrared synchrotron spectral index in Cas A is complicated
in that in {\it J}, {\it H}, and IRAC channel 2, most regions of bright
synchrotron emission are also bright in emission lines. However, a number of isolated
continuum knots are line free. Fortuitously, one of the brightest of these (``knot N''
from W99, marked with a circle in Figure 2), 
is only 13\arcsec\ from Cas A's expansion center
(Thorstensen et al. 2001)
making this the closest feature to the projected SN observed from the optical through 
mid-IR. Thus, a measurement of \AV\ at this location provides the strongest constraint
on the actual reddening of Cas A's SN.

In Figure 4, we plot the $1 - 5$\um\ flux densities for a circular aperture of
radius 10\arcsec\ centered on knot N, with the background (determined from an
annulus of inner radius 10\arcsec\ and outer radius 20\arcsec) subtracted. 
For the ground-based
$JHK_{s}$\ bands, we convert 2MASS magnitudes to Jy using the flux zero-points
from the 2MASS documentation (Cutri et al. 2006). 
The error bars reflect the Poisson noise from
the source and sky, and do not include any systematic error due to large-scale flat-field
error or non-zero color terms in the flux calibration. We estimate these systematics
to be sub-dominant. For the IRAC bands, flux density is a product of the post-BCD
processing. The Poisson errors are negligible. However, due to the extended wings in
the {\it Spitzer}/IRAC point-spread function, there is substantial systematic error
in the flux densities. We correct the fluxes from the images down by 2\% (ch1)
and 5\% (ch2), and assume a conservative 20\% error in the absolute value
\footnote{http://ssc.spitzer.caltech.edu/irac/calib/extcal/index.html}.

In order to estimate \AV, we fit our data with a reddened power law 
and used a Markov Chain
Monte Carlo (MCMC) technique to sample the three-dimensional parameter space
of the fit.
In Figure 5, we plot the likelihood contours
in the $A_{V} - \alpha$\ projection, marginalizing 
(i.e. integrating the likelihood function)
over the normalization parameter. With flat priors on the parameters,
there is a strong degeneracy between the extinction and spectral index,
as higher \AV\ compensates for flatter spectra. This leads to only loose
constraints on reddening. Indeed, the distribution of \AV\ marginalized over both
$\alpha$\ and $k$\ is quite broad and, without further constraints,
we estimate $A_{V} = 7.9 \pm 1.9$.

The data and flat priors alone clearly allow non-physical values of the spectral index,
and improbable reddenings. In order to limit the range of likelihood
space available to the fit, it is clearly desirable to apply more stringent
priors to the parameters. While one could choose somewhat arbitrary allowed ranges
on the parameters (top-hat priors), we choose instead to apply a physically motivated prior
on the spectral index, from the millimeter-wave radio observations. W09
report $\alpha = 0.80$\ for knot N. While they do not quote an error specific to
this knot, they claim their general accuracies to be $\pm 0.02-0.05$. We ran a second
set of Markov chains with a Gaussian prior with $\sigma_{\alpha} = 0.05$. This shrinks the
likelihood contours significantly, and tightly constrains the marginal
probability distribution of the reddening (Figure 5).
For this prior, we find $A_{V} = 6.2 \pm 0.6$.

Given the evidence for departures from a power-law in the broadband synchrotron
spectrum of Cas A discussed previously,
the actual uncertainty in the spectral index may indeed be greater than the
statistical error quoted by W09. The concave-up spectra preferred by both
J03 and R03 flattens the power-law, and would translate to a lower index in
our model. Inspection of our likelihood contours shows that this would
drive our fit to higher \AV. (Indeed, with the flat priors, our data prefers
a flatter $\alpha$\ and more extinction.)
One could construct more exotic priors that preferentially open parameter space
towards lower spectral indices, but absent a strong theoretical guide of their
functional form, and given the relatively paucity of data points, we rapidly approach
art over science. Shorter wavelength data, whose extinction increase faster than
the intrinsic spectrum drops, may have power to constrain more tightly either the spectral
index or any significant departure from a power-law. Imaging of Cas A's synchrotron
emission should be feasible in the red end of the optical with moderate-sized telescopes.
Ideally, one would like a $0.8-2.5$\um\ dispersed spectrum of the emission, though 
this would require a significant investment of time on a large aperture facility.

\begin{figure}
\plotone{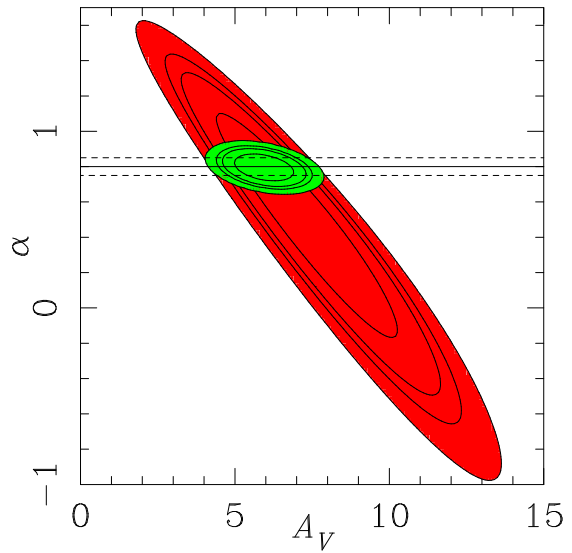}
\caption{Likelihood contours for Cas A knot N for flat priors (red) and 
a Gaussian prior on the spectral index (green). The contours correspond to
68\%, 90\%, 95\%, and 99\%\ confidence for each set of priors. The horizontal
line is the millimeter spectral index from Wright et al. (1999, solid), and their
$1\sigma$\ errors (dashed).}
\end{figure}


\section{Discussion}
\subsection{Cas A's Supernova at Maximum Light}

Accepting our new extinction measurement from synchrotron knot N as the most likely
\AV\ toward Cas A's nascent supernova, we now seek to constrain a range of possible 
intrinsic peak luminosities, and therefore, $^{56}$Ni yields.
Following Young et al. (2006),

\begin{equation}
L_{peak} = M_{Ni}\, \Theta(t_{peak})\, \Lambda(t_{peak})
\end{equation}

\noindent where $M_{Ni}$ is the ejected mass of $^{56}$Ni, $\Theta(t)$\ is the 
instantaneous 
energy decay rate of $^{56}$Ni and $^{56}$Co, and $\Lambda(t)$\ is the efficiency
with which that energy is deposited in the supernova gas. $\Theta(t)$\ is defined as

\begin{equation}
\Theta(t) = \frac{N_{A}}{56} \Bigl[ \frac{E_{Ni}}{\tau_{Ni}} e^{-t/\tau_{Ni}} 
+ \frac{E_{Co}}{\tau_{Co} - \tau_{Ni}} \Bigl( e^{-t/\tau_{Co}} - e^{-t/\tau_{Ni}} \Bigr) \Bigr] .
\end{equation}

Here, $N_{A}$\ is Avogadro's number, while $E$\ and $\tau$ are the $\gamma$-ray energy
and mean decay lifetime of their respective nuclei (1.73 MeV and $7.6 \times 10^{5}$\, s for
nickel; 3.69 MeV and $9.6 \times 10^{6}$\, s for cobalt). We choose
$t_{peak} = 20.7$\ days (the rise time for the SNIIb 2008ax, Pastorello et al. 2008),
and $\Lambda(t_{peak}) = 0.95$\ (the value for the most likely Cas A progenitor
from Young et al., determined with $\gamma$-ray radiative transfer calculations).

First, we estimate the minimum possible luminosity. Willingale et al. (2003) inventory
$0.058\,M_{\sun}$ (no error bar reported) 
of X-ray emitting iron in Cas A from X-ray observations,
and conclude that most of this material is ejecta.
Assuming plausible isotopic ratios, the vast majority of this material is $^{56}$Fe,
the stable daughter product of $^{56}$Ni. 
If we allow Willingale's iron mass to be the sum total of $^{56}$Ni\ ejected, 
the supernova's absolute magnitude would be $M_{V} = -16.5$.
At a distance of $3.4^{+0.3}_{-0.1}$kpc, with $A_{V} = 6.2\, \pm 0.6$, 
this translates to a peak
apparent visual magnitude of $m_{V} = 2.3\, \pm 0.7$. (We have included a 25\%\ error
in the X-ray emitting iron mass.) 

Recently, Krause et al. (2008) observed the actual optical peak of Cas A's supernova
through a light echo, and definitively classified it as a rare Type IIb. They use
the nearby SNIIb SN1993J, which had a peak $M_{V} = -17.5$\ (Richmond et al. 1994)
as a template, assumed the maximum reddening ($A_{V} = 8$) inferred from the
radio, and derived a peak visual magnitude of $m_{V} = 3.2$. Our 
study indicates the extinction is considerably less, with concomitant increase in
apparent brightness. There is, however, significant variation in the peak luminosity of the
peculiar SNIIb class, with SN1996cb peaking at $M_{V} \approx -16.3$\ 
(Qiu et al. 1999), so some care
must be applied in treating the SNIIb as a homogenous population.

Clearly, Flamsteed's apparent observation of 3 Cas was not Cas A's SN at maximum light,
though it may have been the supernova in decline. 
Krause et al.  note that their light echo 
grew fainter by a factor of 18 in 140 days and that this $\sim 3$ magnitude fading matches
a similar decline in lightcurve of 1993J. Morgan (2008) explore the observability of
the putative SN1680 with new atmospheric
radiative transfer calculations and a detailed discussion of the timing within the year.
We summarize his argument here, with our new extinction result.
Given Flamsteed's 16 August 1680 $6^{m}$\
observation, and our assumed peak apparent magnitude of $m \sim 1-3$, SN1680's peak
would have occurred in February-April. During this time, Cas A's position would have
transitted during daylight, and would
not have been visible at maximum altitude by several magnitudes, given any of our
estimates of its apparent magnitude. However, with Cassiopeia's circumpolar
location in the northern sky, SN1680 would still have been quite high in both the
morning and evening twilight. If Cas A's supernova were on the
bright end of our estimates, while it may not have been visible during daylight like
other historical supernovae, it would briefly have been one of the 20 brightest stars 
in the northern sky, and likely would have been reported. Conversely, if its peak magnitude
were towards the fainter end of our range, it may have been just faint enough to
escape notice.

\begin{figure}[t!]
\plotone{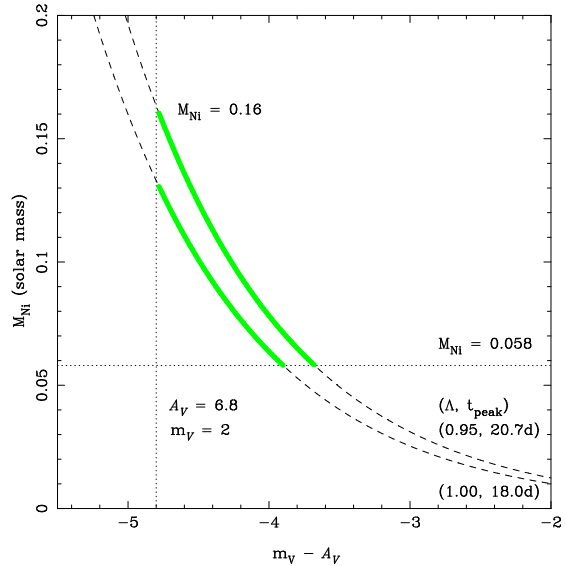}
\caption{$^{56}$Ni mass for two sets of assumptions for the supernova rise time and 
efficiency of $\gamma$-ray energy deposition. The horizontal dotted curve denotes
the X-ray emitting iron mass, while the vertical dotted line represents the brightest
plausible optical transient that might be missed ($m_{V} = 2$), and our best fit
$A_{V} + 1\sigma$. The green solid line marks the allowed range of $^{56}$Ni mass.}
\end{figure}

\subsection{$^{56}$Ni and $^{44}$Ti}

Given our probable range of luminosity for the supernova, we now seek to bound the total mass
of ejected $^{56}$Ni. The detection of $0.058\, M_{\sun}$\ of iron in the
X-ray sets a hard floor. It is more difficult to derive a stringent upper limit, given the 
lack of a rigorous detection threshold. A $0^{m}$\ supernova would have garnered
a great deal of contemporary attention; clearly Cas A's supernova did not. Would a $2^{m}$\ or
$3^{m}$\ transient escape notice? Absent a clear answer, with the definition of
the distance modulus and inverting equation (1) 

\begin{equation}
M_{V} =  m_{v} - A_{V} - \mu \propto \log_{10} M_{Ni}
\end{equation}

\noindent we combine our ignorance into the $m_{V} - A_{V}$.  In Figure 6,
we plot this difference against the calculated ejected mass of nickel. The lower limit
on $M_{Ni}$, combined with our best fit \AV\ implies $m_{V} < 2.4$. This is 
uncomfortably bright, but not an entirely unreasonable magnitude for an object that
may have gone unnoticed in the twilight.
If, instead we assume at $1\sigma$\ increase over our
best \AV\ and assume that any transient with $m < 2$\ would merit at least a historical
footnote, we place an upper limit on $M_{Ni} < 0.16 M_{\sun}$. We regard this 
as a conservative estimate. Plausibly shorter rise times and larger energy deposition
efficiencies can lower this limit by $\sim20\%$.

Gamma-ray observations have detected the nuclear decay lines of 
 $^{44}$Ti daughter nuclei $^{44}$Ca and $^{44}$Sc (Iyudin et al. 1994, Vink et al. 2001)
The latest $\gamma$-ray observations imply an
initially synthesized $^{44}$Ti mass of $1.6^{+0.6}_{-0.3} \times 10^{-4}\, M_{\sun}$\
(Renaud et al. 2006),  for a $^{44}$Ti/$^{56}$Ni $= 0.8 - 3.8 \times 10^{-3}$.
This high mass and ratio is problematic for spherically symmetric explosion
calculations (Timmes et al. 1996), though in multi-dimensional asymmetric 
simulations some mix of differentially enhanced $\alpha$-rich
freezeout (Nagataki et al. 1998) and differential fall-back (Young et al. 2006)
appear capable of reproducing a range of yields, some consistent with the observed
value. Notably, the combination of the new {\it INTEGRAL}\ titanium measurements and
our tighter bound on the ejected $^{56}$Ni mass shrink considerably the allowed
abundance ratio space considered by Young et al (2006), and may provide important
constraints on advanced supernova explosion calculations. Finally, we note that our
radioactive Ti/Ni bound is consistent with the solar $^{44}$Ca/$^{56}$Fe ratio
($1.5 \times 10^{-3}$, Anders \& Grevesse 1989)
the stable isotopes of the considered decay chains. While is it dangerous
to extrapolate from one observation, if this result is generally valid,
Cas A-like supernovae may lessen
the need for the sub-Chandrasekhar mass SNIa required by Timmes et al. (2006) to match the
solar abundance.

Finally, we note an alternative explanation for the apparent faintness of Cas A's
supernova. While nearly all long duration gamma-ray bursts show late-time supernova
light curves, a small number do not
(e.g.s. GRB060606 and GRB060615, Fynbo et al. 2006).
Calculations by Fryer et al. (2007) show that sufficiently
delayed explosions (after core collapse) produce a large mass of material on
gravitationally bound ballistic trajectories, which
falls back onto the compact remnant (a black hole for GRBs) on a timescale
of minutes after the explosion. A correspondingly small amount
of nickel escapes into the remnant. In these cases, the lightcurve is powered not by
radioactive decay, but by energy deposited by the blast wave. 
These supernova are fainter and decline faster. In principle, similar such delayed
supernovae could exist independently of GRBs, and their outbursts
would be easier to hide. This idea is attractive in that the spectra of 
a number of the oxygen-rich supernova remnants lack iron,
or even oxygen-burning ashes such as sulfur, argon, and calcium (e.g. N132D
and E0102, Blair et al. 2000),
which may well have fallen back onto the compact remnant. However,
given the significant mass of iron detected in the X-ray spectrum of Cas A, its supernova
was unlikely to have been such an event, so interstellar extinction remains the
likely culprit in its apparent faintness.


\section{Conclusion}

We have presented new extinction measurements for Cassiopeia A that show the reddening
to be more variable and of higher magnitude than had been derived from previous optical
measurements, though our estimates of \AV\ are consistent with previous radio observations.
The \feii\ lines should be promising probes of reddening for regions of moderate extinction,
though their use is still hampered by uncertainty in the intrinsic emissivity ratio 
of the two lines. Further atomic structure calculations, or more appropriately, observations
of unabsorbed emission line sources are necessary to yield the full utility of these features.
Nevertheless, the observed ratio is sufficient to show the east-west reddening gradient
previously inferred in the radio, as well as small scale point-to-point variations.

We have also shown that estimation of \AV\ from the infrared synchrotron index 
combined with constraints from the radio yields
statistical errors comparable to or better than those from the emission lines.
Fortuitously, a bright synchrotron knot free from emission-line contamination is located 
just 13\arcsec\ away from the derived expansion center of the remnant, and thus provides
the strongest constraint on \AV\ toward Cas A's nascent supernova. Our new 
reddening, taken with the lack of widespread reportage of the supernova and with an iron
mass inventory from the X-ray, bounds the ejected mass of $^{56}$Ni to be 
$0.058 - 0.16 M_{\sun}$. This nickel abundance and measurement of the $^{44}$Ti yield
from $\gamma$-radiation both provide constraints on possible asymmetry in the supernova
explosion, and imply a $^{44}$Ca/$^{56}$Fe ratio consistent with solar, which
may have consequences for Galactic chemical evolutionary models.

{\it Facilities:} \facility{Bok (PISCES imager)}, \facility{Mayall (FLAMINGOS spectrograph)}, \facility{Spitzer (IRAC imager)}


\newpage

\begin{thebibliography}{}
\bibitem[Allen et al.(1997)]{1997ApJ...487L..97A} Allen, G.~E., et al.\ 1997, \apjl, 487, L97 
\bibitem[Anders \& Grevesse(1989)]{1989GeCoA..53..197A} Anders, E., \& Grevesse, N.\ 1989, \gca, 53, 197 
\bibitem[Ashworth(1980)]{1980JHA....11....1A} Ashworth, W.~B., Jr.\ 1980, Journal for the History of Astronomy, 11, 1 
\bibitem[Blair et al.(2000)]{2000ApJ...537..667B} Blair, W.~P., et al.\ 2000, \apj, 537, 667 
\bibitem[Bohlin(2007)]{2007ASPC..364..315B} Bohlin, R.~C.\ 2007, The Future 
of Photometric, Spectrophotometric and Polarimetric Standardization, 364, 
315 
\bibitem[Cardelli et al.(1989)]{1989ApJ...345..245C} Cardelli, J.~A., 
Clayton, G.~C., \& Mathis, J.~S.\ 1989, \apj, 345, 245 
\bibitem[Cutri et al. (2006)]{Cutri2MASSExplan} Cutri, R.~M., et al.\ 2006, Explanatory Supplement to the 2MASS All Sky Data Release and Extended Mission Products, http://www.ipac.caltech.edu/2mass/releases/allsky/doc/explsup.html
\bibitem[Eichler(1984)]{1984ApJ...277..429E} Eichler, D.\ 1984, \apj, 277, 429 
\bibitem[Ennis et al.(2006)]{2006ApJ...652..376E} Ennis, J.~A., Rudnick, 
L., Reach, W.~T., Smith, J.~D., Rho, J., DeLaney, T., Gomez, H., 
\& Kozasa, T.\ 2006, \apj, 652, 376 
\bibitem[Fesen et al.(2006)]{2006ApJ...636..848F} Fesen, R.~A., Pavlov, G.~G., \& Sanwal, D.\ 2006, \apj, 636, 848 
\bibitem[Fynbo et al.(2006)]{2006Natur.444.1047F} Fynbo, J.~P.~U., et al.\ 2006, \nat, 444, 1047 \bibitem[Fryer et al.(2007)]{2007ApJ...662L..55F} Fryer, C.~L., Hungerford, A.~L., \& Young, P.~A.\ 2007, \apjl, 662, L55 
\bibitem[Gredel \& Reipurth(1993)]{1993ApJ...407L..29G} Gredel, R., \& Reipurth, B.\ 1993, \apjl, 407, L29 
\bibitem[Gerardy \& Fesen(2001)]{2001AJ....121.2781G} Gerardy, C.~L., \& Fesen, R.~A.\ 2001, \aj, 121, 2781 
\bibitem[Gray \& Corbally(1994)]{1994AJ....107..742G} Gray, R.~O., \& Corbally, C.~J.\ 1994, \aj, 107, 742 

\bibitem[Hartigan et al.(2004)]{2004ApJ...614L..69H} Hartigan, P., Raymond, 
J., \& Pierson, R.\ 2004, \apjl, 614, L69 
\bibitem[Hughes et al.(2000)]{2000ApJ...528L.109H} Hughes, J.~P., Rakowski, 
C.~E., Burrows, D.~N., \& Slane, P.~O.\ 2000, \apjl, 528, L109 
\bibitem[Hurford \& Fesen(1996)]{1996ApJ...469..246H} Hurford, A.~P., \& Fesen, R.~A.\ 1996, \apj, 469, 246 
\bibitem[Iyudin et al.(1994)]{1994A&A...284L...1I} Iyudin, A.~F., et al.\ 1994, \aap, 284, L1 
\bibitem[Jones et al.(2003)]{2003ApJ...587..227J} Jones, T.~J., Rudnick, L., DeLaney, T., \& Bowden, J.\ 2003, \apj, 587, 227 
\bibitem[Kamper(1980)]{1980Obs...100....3K} Kamper, K.~W.\ 1980, The Observatory, 100, 3 
\bibitem[Keenan et al.(1993)]{Keenan93} Keenan, F.~P., Hibbert, A., Ojha, P.~C., Conlon, E.~S.\ 1993, \physscr, 48, 129
\bibitem[Keenan et al.(1996)]{1996MNRAS.281.1073K} Keenan, F.~P., Aller, L.~H., Bell, K.~L., Hyung, S., McKenna, F.~C., \& Ramsbottom, C.~A.\ 1996, \mnras, 281, 1073 
\bibitem[Keohane et al.(1996)]{1996ApJ...466..309K} Keohane, J.~W., Rudnick, L., \& Anderson, M.~C.\ 1996, \apj, 466, 309 
\bibitem[Klazebrook \& Economou (1997)]{GE97}Glazebrook, K., \& Economou, F. 1997, Dr. Dobb's Journal, 9719, 45
\bibitem[Krause et al.(2008)]{2008Sci...320.1195K} Krause, O., Birkmann, S.~M., Usuda, T., Hattori, T., Goto, M., Rieke, G.~H., \& Misselt, K.~A.\ 2008, Science, 320, 1195 
\bibitem[Lamers et al.(1983)]{1983A&A...128..299L} Lamers, H.~J.~G.~L.~M., de Groot, M., \& Cassatella, A.\ 1983, \aap, 128, 299 
\bibitem[Laming(2001)]{2001ApJ...546.1149L} Laming, J.~M.\ 2001, \apj, 546, 1149 
\bibitem[McCarthy et al.(2001)]{2001PASP..113..353M} McCarthy, D.~W., Jr., 
Ge, J., Hinz, J.~L., Finn, R.~A., \& de Jong, R.~S.\ 2001, \pasp, 113, 353 
\bibitem[Morgan(2008)]{2008Obs...128...80M} Morgan, J.~A.\ 2008, The Observatory, 128, 80 
\bibitem[Nagataki et al.(1998)]{1998ApJ...492L..45N} Nagataki, S., Hashimoto, M.-A., Sato, K., Yamada, S., \& Mochizuki, Y.~S.\ 1998, \apjl, 492, L45 
\bibitem[Nisini et al.(2005)]{2005A&A...441..159N} Nisini, B., Bacciotti, F., Giannini, T., Massi, F., Eisl{\"o}ffel, J., Podio, L., \& Ray, T.~P.\ 2005, \aap, 441, 159 
\bibitem[Nussbaumer \& Storey(1988)]{1988A&A...193..327N} Nussbaumer, H., \& Storey, P.~J.\ 1988, \aap, 193, 327 
\bibitem[Pastorello et al.(2008)]{2008MNRAS.389..955P} Pastorello, A., et 
al.\ 2008, \mnras, 389, 955 
\bibitem[Qiu et al.(1999)]{1999AJ....117..736Q} Qiu, Y., Li, W., Qiao, Q., \& Hu, J.\ 1999, \aj, 117, 736 
\bibitem[Quinet et al.(1996)]{1996A&AS..120..361Q} Quinet, P., Le Dourneuf, M., \& Zeippen, C.~J.\ 1996, \aaps, 120, 361 
\bibitem[Reynoso \& Goss(2002)]{2002ApJ...575..871R} Reynoso, E.~M., \& Goss, W.~M.\ 2002, \apj, 575, 871 
\bibitem[Rho et al.(2003)]{2003ApJ...592..299R} Rho, J., Reynolds, S.~P., 
Reach, W.~T., Jarrett, T.~H., Allen, G.~E., \& Wilson, J.~C.\ 2003, \apj, 592, 299 
\bibitem[Reed et al.(1995)]{1995ApJ...440..706R} Reed, J.~E., Hester, J.~J., Fabian, A.~C., \& Winkler, P.~F.\ 1995, \apj, 440, 706 
\bibitem[Reynolds(1998)]{1998ApJ...493..375R} Reynolds, S.~P.\ 1998, \apj, 493, 375 
\bibitem[Richmond et al.(1994)]{1994AJ....107.1022R} Richmond, M.~W., Treffers, R.~R., Filippenko, A.~V., Paik, Y., Leibundgut, B., Schulman, E., \& Cox, C.~V.\ 1994, \aj, 107, 1022 
\bibitem[Rousselot et al.(2000)]{2000A&A...354.1134R} Rousselot, P., Lidman, C., Cuby, J.-G., Moreels, G., \& Monnet, G.\ 2000, \aap, 354, 1134 
\bibitem[Searle(1971)]{1971ApJ...168...41S} Searle, L.\ 1971, \apj, 168, 41 
\bibitem[Skrutskie et al.(2006)]{2006AJ....131.1163S} Skrutskie, M.~F., et 
al.\ 2006, \aj, 131, 1163 
\bibitem[Smith \& Hartigan(2006)]{2006ApJ...638.1045S} Smith, N., \& Hartigan, P.\ 2006, \apj, 638, 1045 
\bibitem[Timmes et al.(1996)]{1996ApJ...464..332T} Timmes, F.~X., Woosley, 
S.~E., Hartmann, D.~H., \& Hoffman, R.~D.\ 1996, \apj, 464, 332 
\bibitem[Thorstensen et al.(2001)]{2001AJ....122..297T} Thorstensen, J.~R., Fesen, R.~A., \& van den Bergh, S.\ 2001, \aj, 122, 297 
\bibitem[Troland et al.(1985)]{1985ApJ...298..808T} Troland, T.~H., 
Crutcher, R.~M., \& Heiles, C.\ 1985, \apj, 298, 808 
\bibitem[Vink et al.(2001)]{2001ApJ...560L..79V} Vink, J., Laming, J.~M., 
Kaastra, J.~S., Bleeker, J.~A.~M., Bloemen, H., \& Oberlack, U.\ 2001, \apjl, 560, L79 
\bibitem[Willingale et al.(2003)]{2003A&A...398.1021W} Willingale, R., Bleeker, J.~A.~M., van der Heyden, K.~J., \& Kaastra, J.~S.\ 2003, \aap, 398, 1021 
\bibitem[Wright et al.(1999)]{1999ApJ...518..284W} Wright, M., Dickel, J., Koralesky, B., \& Rudnick, L.\ 1999, \apj, 518, 284 
\bibitem[Young et al.(2006)]{2006ApJ...640..891Y} Young, P.~A., et al.\ 2006, \apj, 640, 891 


\end{thebibliography}
\end{document}